\documentstyle[preprint,aps]{revtex}

\tightenlines

\newcommand{\lsim}[1]{
\setlength{\unitlength}{12pt}
\begin{picture}(1.4,1.)
\put(.7,-0.3){\makebox(0.0,1.)[t]{$<$}}
\put(.7,-0.3){\makebox(0.0,1.)[b]{$\sim$}}
\end{picture}#1}
\newcommand{\gsim}[2]{
\setlength{\unitlength}{12pt}
\begin{picture}(1.4,1.)
\put(.7,-0.3){\makebox(0.0,1.)[t]{$>$}}
\put(.7,-0.3){\makebox(0.0,1.)[b]{$\sim$}}
\end{picture}#2}
\begin{document}
\draft

\title{Nonstandard neutrino properties and the high-energy cosmic neutrino
flux}

\author{R. Horvat \\
  ``Rudjer Bo\v skovi\' c'' Institute, P.O.Box 1016, 10001 Zagreb,
Croatia}

\maketitle

\begin{abstract}
Recently, it has been shown that cosmic neutrino flux at the GRB/AGN source,
$F_{\nu_e }^0 : F_{\nu_{\mu }}^0 : F_{\nu_{\tau }}^0 \simeq 1 :2 :0 $ ,
inevitably oscillates (in the three neutrino framework) to 1 : 1: 1, 
irrespective of the mixing angle relevant to the solar data. It has also been 
pointed out that the intrinsic flux of the cosmic high-energy neutrinos may 
not have the standard ratio, in which case the cosmic neutrino flux in the 
far distance should be dependent of the mixing angle. For neutrinos with 
nonzero electromagnetic properties, we show in the latter case that matter 
effects on oscillations of high-energy cosmic neutrinos can substantially 
modify the vacuum oscillation probability, if the mass-squared difference is
$ \lsim 10^{-10} \;\mbox{\rm eV}^2 $.  
\end{abstract}

\newpage

Recently, AJY \cite{1} have made a detailed numerical analysis to show that
the final flux of high-energy cosmic neutrinos turns out to be almost
equally distributed amongst the three flavors, provided that neutrinos are
produced by cosmologically distant objects like Gamma Ray Bursters (GRBs)
and Active Galactic Nuclei (AGN). When they had extended their analysis to
four neutrinos, they found a similar result. Later on, it was shown
analytically by AOA \cite{2} that the above result follows directly from a
quasi bi-maximal mixing matrix \cite{3}, implied by the oscillation solution
of the atmospheric neutrino problem. In addition, they found that the
prediction is independent of the mixing angle responsible for the solar
neutrino anomaly.

The modification of the properties of high-energy neutrino fluxes in
\cite{1,2} was due to vacuum oscillations on the way from cosmologically
distant sources to the Earth. The study of oscillation effects on
high-energy neutrino flux can thus offer the possibility to probe neutrino
mixing and distinguish between different mass schemes \cite{1}. 

The matter effects on oscillations of high-energy cosmic neutrinos have been
studied recently by Lunardini and Smirnov in \cite{4}. They found, by
considering the relic neutrino background as the only relevant matter
background, that substantial effect can be produced, provided the background
has large CP asymmetry. In three-neutrino schemes, however, no matter effects
appear since the final cosmic flux does not depend on the mixing angle
\cite{1,2}. 

To study matter effects in three-neutrino schemes one has therefore to rely
on a hypothesis that the intrinsic flux of  cosmic neutrinos  does not
have the standard ratio, $F_{\nu_e }^0 : F_{\nu_{\mu }}^0 : F_{\nu_{\tau
}}^0
\simeq 1 :2 :0 $. As stressed in \cite{5}, this indeed may be the case, if
some of the muons lose their energy in the relatively intense magnetic field
in the vicinity of the source. In what follows, we will explore the
consequences of a generic neutrino flux at the source, $F_{\nu_e }^0 :
F_{\nu_{\mu }}^0 : F_{\nu_{\tau }}^0 \simeq \frac{\lambda }{3} : 1-
\frac{\lambda }{3} : 0 $, where $\lambda $ is taken from the interval $0
\leq \lambda \leq 1 $, with $\lambda = 1 $ corresponding to the standard
case. In the above flux ratio, in the symbol for neutrinos both neutrinos
and antineutrinos are counted.

In the present paper, we show that for neutrinos with nonzero
electromagnetic properties (magnetic moment/charge radius), the final
neutrino flux can be substantially modified with respect to the vacuum
oscillation case, for the allowed values of nonstandard neutrino intrinsic
parameters. In addition, large matter effects are possible even for a
perfect CP symmetry of the relic neutrino background. There is a crucial
observation which makes interesting to study the conversion effects of
high-energy neutrinos due to the refraction in the relic neutrino sea, when
neutrinos possess nonzero electromagnetic properties. It is known that the
refractive potential $V$ for a neutrino 
due to neutrino-relic neutrino interaction is tiny even for large
CP asymmetries $( \lsim \;\mbox{\rm 10}^{-30} \;\mbox{\rm eV})$. However,
the relevant quantity for neutrino oscillation in matter appears to be the
induced mass squared for a neutrino, where the potential is multiplied by
the energy $E_{\nu }$, which, in the high-energy case under consideration,
lies in the range $\mbox{\rm 10}^{15} \;\mbox{\rm eV} \lsim E_{\nu } \lsim
\mbox{\rm
10}^{22} \;\mbox{\rm eV}$. Hence, we see that matter effects for these
high-energy neutrinos can be important even for tiny $V$. For
neutrinos with nonzero electromagnetic properties, we show below that the
refraction potential itself is proportional to $E_{\nu }$, leading to the
induced mass squared proportional $E_{\nu }^2 $. So, one expects matter
effects to be enhanced relative to the standard case, where the induced mass
squared is proportional to $E_{\nu }$.

Let us begin with the mixing matrix implied by the oscillation solution of
the atmospheric neutrino problem in the three flavor framework:  
\begin{equation}
  U = \left( \begin{array}{ccc}
         c_{\theta } & \;\;\; s_{\theta } & \;\;\; 0 \\
         \frac{-s_{\theta }}{\sqrt{2}} & \;\;\; \frac{c_{\theta
}}{\sqrt{2}} & \;\;\; \frac{1}{\sqrt{2}} \\
        \frac{s_{\theta }}{\sqrt{2}} & \;\;\; \frac{-c_{\theta
}}{\sqrt{2}} & \;\;\; \frac{1}{\sqrt{2}}
    \end{array}     \right) \;,
\label{form1}
\end{equation}
where we take $\theta $ to be the mixing angle relevant for the solar data
\cite{3}, and $c_{\theta } \equiv \cos{\theta }$, $s_{\theta } \equiv
\sin{\theta }$. The oscillated flux of high-energy neutrinos, $F_{l}^D $, 
measured at terrestial detectors is given by
\begin{equation}
F_{l}^D = \sum_{l^{'}} P_{l l^{'}} F_{l^{'}}^0 \;.
\label{form2}
\end{equation}
For cases where the production positions of high-energy neutrinos extend
over a distance much larger than wavelength, one usually deals with a
classical probability, obtained as the product of the magnitudes squared of
the matrix $U$, instead of the magnitude squared of the product. Hence, Eq.
(1) yields:
\begin{equation}
   P = \left( \begin{array}{ccc}
 s_{\theta }^4 + c_{\theta }^4 & \;\;\; c_{\theta }^{2}s_{\theta }^{2} &
\;\;\; c_{\theta }^{2}s_{\theta }^{2} \\
c_{\theta }^{2}s_{\theta }^{2} & \;\;\; \frac{1}{4}[1 + s_{\theta }^4 +
c_{\theta }^4 ] & \;\;\; \frac{1}{4}[1 + s_{\theta }^4 + c_{\theta }^4 ] \\
c_{\theta }^{2}s_{\theta }^{2} & \;\;\; \frac{1}{4}[1 + s_{\theta }^4 +
c_{\theta }^4 ] & \;\;\; \frac{1}{4}[1 + s_{\theta }^4 + c_{\theta }^4 ]
\end{array}     \right) \;.
\label{form3}
\end{equation}
Using (3) in (2), yields the prediction \cite{2}, $F_{\nu_e }^D : F_{\nu_{\mu
}}^D : F_{\nu_{\tau }}^D = 1 : 1 : 1 $, a result obtained also in \cite{1}
through a detailed numerical analysis. An interesting feature of the above 
result is the independence of the final ratio from the mixing angle $\theta $.
Hence, even if $\theta $ is largely affected by matter effects, they remain
unobservable in the final flux.

Here, we consider a generic source ratio, $F_{\nu_e }^0 :
F_{\nu_{\mu }}^0 : F_{\nu_{\tau }}^0 \simeq \frac{\lambda }{3} : 1-
\frac{\lambda }{3} : 0 $, with $0
\leq \lambda \leq 1 $, consistent with a possibility that muons lose their
energy in a magnetic field. The standard ratio is obtained for $\lambda = 1
$. Now, the final flux explicitly depends on the mixing angle $\theta $ as
\begin{equation}
\left( \begin{array}{c} F_{\nu_e }^D \\[1mm] F_{\nu_{\mu }}^D \\[1mm] F_{\nu_{\tau
}}^D \end{array} \right) = \left( \begin{array}{ccc} \frac{\lambda }{3} +
(1-{\lambda })c_{\theta }^{2}s_{\theta }^{2} \\[1mm] -\frac{\lambda }{6} +
\frac{1}{2}
+(1/2)({\lambda }-1)c_{\theta }^{2}s_{\theta }^{2} \\[1mm] -\frac{\lambda }{6} +   
\frac{1}{2}
+(1/2)({\lambda }-1)c_{\theta }^{2}s_{\theta }^{2} \end{array}
\right) C \;,
\label{form4}
\end{equation}
where $C$ is some common flux. The effective mixing angle in matter 
${\theta }_m $ for the above scenario reads,
\begin{equation}
s_{2 {\theta }_m } = \frac{\Delta s_{2 {\theta }}}{[(A - \Delta c_{2 {\theta
}})^2 + (\Delta s_{2 {\theta }})^2]^{1/2}} \;\;\;,
\label{form5}
\end{equation}
where $A $ is the induced mass squared for the neutrino (to be specified
accurately below), and $\Delta $ is the mass squared difference relevant for
the solar neutrino problem. In what follows we will always adopt a
criterion,
\begin{equation}
\mid A \mid \; \gsim \; \Delta \;,
\label{form6}
\end{equation}
to quantify the importance of matter effects for high-energy neutrinos
crossing cosmological distances in the universe.
 
For neutrino parameters and energies relevant for the flux discussion (see
below), the
absorption effects are always negligible, and therefore the expression (5)
for the matter angle is always valid. Let us stress that matter effects of
the sources, for a range of neutrino masses and mixing relevant for the solar
neutrino problem, can be neglected \cite{6}. On the other hand, for
extremely high-energy neutrinos $(E_{\nu } \gsim \;\mbox{\rm 10}^{21}
\;\mbox{\rm eV})$ and eV-mass neutrino background, resonant neutrino
conversion is possible in dark matter
halos \cite{7}. This is due to the fact that the halo background is no longer
flavor symmetric when neutrino masses are hierarchal \cite{7}.

Let us first assume that neutrinos are of the Dirac type, and consider
neutrino oscillations for cosmic neutrinos with anomalous magnetic moments
\footnote{The nonstandard properties are assumed for source as well as for
background neutrinos.}. The coupling of the electromagnetic field to neutrino
magnetic dipole moments is given by (with family indices suppressed)
\begin{equation}
L = - \frac{1}{2} \mu_{\nu } \bar{\nu}_R  \sigma_{\alpha \beta } \nu_L
F^{\alpha \beta } \; \; \; + h.c. \;. 
\label{form7}
\end{equation}

For neutrinos with $\mu_{\nu } \neq 0 $, the $\nu_R $ background and the
electromagnetic background will affect $\nu_L $'s propagation through the
coherent forward amplitude in the lowest order. The refractive potential can
be easily calculated via the real part of the neutrino thermal self-energy
due to photon-magnetic moment interaction \cite{8}. Let us stress that a 
graph with a cut through the photonic line vanishes because of gauge
invariance. Ignoring the asymmetry term which is a constant, our concern
here will be a CP-symmetric contribution, which, being proportional to the 
incident neutrino energy, can be potentially large. 

The CP-symmetric contribution to the neutrino refractive potential can be
obtained by a slight modification of the results from Ref. \cite{8} as 
\begin{equation}
V = (\pi /12) \alpha {\eta }_{\nu }^2 (E_{\nu }/m_{e}^2 ) T_{\nu_R }^2 \; \; \;,
\label{form8}
\end{equation}
where $\alpha $ is the fine-structure constant, ${\mu }_{\nu } = \eta_{\nu }
\mu_B $, with $\mu_B $  
being the Bohr magneton, and $T_{\nu_R }$ is the temperature of the
$\nu_R $ background. We find for ${\eta }_{\nu } \lsim \mbox{\rm 10}^{-12}$
that $T_{\nu_R } \lsim \mbox{\rm 0.47}\; T_{\nu_L }$ \cite{8}, where $T_{\nu_L } =
(4/11)^{\frac{1}{3}}\;T_{\gamma}$. 
In the following we will consider the two
types of magnetic moment hierarchy, ${\eta }_{{\nu }_e } > {\eta }_{{\nu
}_{\mu}} \simeq {\eta }_{{\nu }_{\tau}}$ and ${\eta }_{{\nu   
}_{\mu}} \simeq {\eta }_{{\nu }_{\tau}} > {\eta }_{{\nu }_e }$. In the former
case $A \simeq 2 E_{{\nu }_e } V_{{\nu }_e }$, while $A \simeq 
-2 E_{{\nu }_{\mu }}
V_{{\nu }_{\mu }}$ in the latter. When matter effects become dominant, $\mid
A \mid \; >> \; \Delta $, $\theta_m \rightarrow \pi/2 $ $(A > 0)$ and
$\theta_m \rightarrow 0 $ $(A < 0)$. Since matter effects enter (4) only
through the combination $c_{\theta }s_{\theta }$, we have in  both cases
that $c_{{\theta }_m } s_{{\theta }_m } \rightarrow 0$
\footnote{We assume that sterile neutrinos do not participate in mixing with
active species.}.
 
Now, we take two-neutrino vacuum oscillations of $\nu_e $ (VO) as a
reference, for which the solar data indicate \cite{9}
\begin{equation}
\begin{array}{l}
\Delta \; = \; 7.5 \times \mbox{\rm 10}^{-11} \; {\mbox{\rm eV}}^2 \\[1mm]  
s_{2 \theta }^2 \; = \; 0.91 \;. 
\end{array} 
\label{form9}
\end{equation}
From (9) we have $c_{\theta }s_{\theta } \simeq 0.476 $, and we therefore 
see that when matter effects dominate, the mixing angle parameter in the
final neutrino flux (4) can (depending on $\lambda $) substantially deviate
from its vacuum value. Applying the constraint (6) to the VO solution (9),
we find that intense fluxes of neutrinos with energies 
\begin{equation}
E_{\nu } \; \gsim \; 9 \times \mbox{\rm 10}^{17} \; \mbox{\rm eV}
\left (\frac{\mbox{\rm 10}^{-12}}{\eta_{\nu }} \right )
\label{form10}
\end{equation}
are affected by matter effects. Since intense fluxes of neutrinos with
energies up to $\mbox{\rm 10}^{21} - \mbox{\rm 10}^{22} \; \mbox{\rm eV}$
are assumed to be produced by cosmologically distant sources like
GRBs/AGN, the limit (10) is relevant even for the most restrictive
astrophysical constraint on neutrino magnetic dipole moments \cite{10}. At
the production site with the redshift $Z$, the bound (10) is relaxed by a
factor $(1 + Z)^{-1}$.

For the homogeneous neutrino background characterized by a single
temperature $T_{\nu } \sim 1.95 \; K$, the energy independent classical
probability (3) with $\theta \rightarrow \theta_m $ is valid if [at the same
time respecting (10)]  
\begin{equation}
L \; >> \; \frac{4 \pi E_{\nu }}{A} = \frac{2 \pi }{V_{\nu }}  \;,
\label{form11}
\end{equation}
where for $L$ we take 500 Mpc as a typical distance of GBRs/AGN. From
(11) we get a constraint,
\begin{equation}
E_{\nu } \; >> \; 1.8 \times \mbox{\rm 10}^{15} \; \mbox{\rm eV} \left
(\frac{\mbox{\rm 10}^{-12}}{\eta_{\nu }} \right )^2 \;,
\label{form11}
\end{equation}
which is consistent with (10) for magnetic moments not  much smaller than
the present astrophysical bounds. If we include the effect of scaling of the
neutrino energy and neutrino temperature with redshift, the constraint (12)
is always less stringent. 

If the universe possesses  the $\nu_R $ asymmetry, 
there is an enhancement of
the relic neutrino density, resulting in a relaxed  bound on $E_{\nu }$ with
respect to (10). In terms of degeneracy parameter $\xi_{{\nu }_R} \equiv
\frac{\kappa_{{\nu }_R }}{T_{{\nu }_R }}$ ($\kappa_{{\nu }_R }$ being the chemical
potential), we obtain
\begin{equation}
E_{\nu } \; >> \; 1.1 \times \mbox{\rm 10}^{17} \; \mbox{\rm eV} \left
(\frac{\mbox{\rm 10}^{-12}}{\eta_{\nu }} \right ) \left (\frac{\mbox{\rm
10}}{\xi_{{\nu }_R }} \right )\;,
\label{form12}                                  
\end{equation}
and a similar bound from the classical probability requirement,
\begin{equation}
E_{\nu } \; >> \; 2.9 \times \mbox{\rm 10}^{13} \; \mbox{\rm eV} \left
(\frac{\mbox{\rm 10}^{-12}}{\eta_{\nu }} \right )^2
\left (\frac{\mbox{\rm
10}}{\xi_{{\nu }_R }} \right )^2 \;.
\label{form13}
\end{equation}
 We should stress here that the BBN light-element
abundance constraints can be satisfied for cosmological models in which
significant neutrino degeneracies \footnote{Due to mixing, at present the
asymmetries in the individual flavors may differ from that at the BBN epoch,
see \cite{4}.},  $\xi_{{\nu }_e } \; \lsim \; {\cal O}(1),
\xi_{{\nu }_{\mu , \tau }} \; \lsim \; {\cal O}(10)$ exist \cite{11}.

For the case of Majorana neutrinos (having off-diagonal magnetic moments),
the only hierarchy among $\eta_{\nu }$'s which mimics the single $\eta_{\nu
}$ dominance introduced earlier for the Dirac case
is $\eta_{{\nu }_{\mu }{\nu }_{\tau }} >> \eta_{{\nu }_e {\nu }_{\mu }}, 
\eta_{{\nu }_e {\nu }_{\tau }}$ $(A < 0)$. This is due to the fact that now
all neutrino flavors are present in the background and that each flavor has
two independent contributions from the thermal self-energy graph.  
It is now straightforward to obtain the lower bound on $E_{\nu }$ where
matter effects start to dominate, since we only need to replace $T_{{\nu }_R
}
\rightarrow T_{{\nu }_L } $ in (8). In this way we obtain 
\begin{equation}
E_{\nu } \; \gsim \; 2 \times \mbox{\rm 10}^{17} \; \mbox{\rm eV}
\left (\frac{\mbox{\rm 10}^{-12}}{\eta_{\nu }} \right )\;,
\label{form14}
\end{equation}
and the corresponding bound from the classical probability requirements now
reads
\begin{equation}
E_{\nu } \; >> \; 3.8 \times \mbox{\rm 10}^{14} \; \mbox{\rm eV} \left
(\frac{\mbox{\rm 10}^{-12}}{\eta_{\nu }} \right )^2 \;.
\label{form15}
\end{equation}

A further static electromagnetic property of the neutrino to be considered
here is its charge radius $<r^2 >$. The presently existing laboratory and
astrophysical limits on the neutrino charge radius imply \cite{12}
\begin{equation}
<r^2 > \; \leq \; \mbox{\rm 10}^{-32} \; \mbox{\rm cm}^{2} \;.
\label{form16}
\end{equation}
From the finite temperature/density graph involving nonvanishing $<r^2 >$,
we find the effective potential for $\nu_L $ as 
\begin{equation}
V \; \simeq \; 0.03 \; <r^2 > E_{\nu } T_{\nu }^4 \;,
\label{form17}                                           
\end{equation}                                           
which is again proportional to $E_{\nu }$. By applying the criterion (5), we
find a  limit, $E_{\nu } \; >> \; 5 \times \mbox{\rm 10}^{25}
\; \mbox{\rm eV}$. Hence, matter effects for cosmic neutrinos originating
from GRBs/AGN and having nonvanishing $<r^2 >$'s are insignificant.

We are now in position to compare the source and the
oscillated flux without and with matter effects. The oscillated flux is
assumed to respect the VO solution as well as the corresponding constraints
[like (10) and (12)]. For $\lambda = 1$ they are the same, whereas 
the biggest difference is obtained for $\lambda = 0$. In the latter case we
have (0.23, 0.385, 0.385) versus (0, 0.5, 0.5).   
 
In conclusion, we have studied matter effects, contributed from the
homogeneous relic neutrino background, on oscillations of high-energy
neutrinos  emitting from cosmologically distant sources like GRBs/AGN and
possessing nonvanishing electomagnetic properties. We assumed the standard
three flavor framework as well as the nonstandard intrinsic neutrino flux
parameterized by a single parameter $\lambda $. For the astrophysical upper
limit on neutrino magnetic moments $(\eta_{\nu } \simeq \mbox{\rm
10}^{-12})$, we found that matter effects can be important in the energy
interval $E_{\nu } \; \geq \; {\cal O}(10^{17})$ eV, 
if the VO solution is chosen. This means that any fair departure of
the parameter $\lambda $ from the value $\lambda = 1$, can lead to a drastic
change of the oscillated flux of high-energy neutrinos.

{\bf Acknowledgments. } The author acknowledges the support of the Croatian
Ministry of Science and Technology under the contract 1 -- 03 -- 068.

\end{document}